\documentstyle[11pt]{article}
\def\bsh{\backslash}
\def\bdt{\dot \beta}
\def\adt{\dot \alpha}

\newfont{\bbbold}{msbm10}

\def\bbC{\mbox{\bbbold C}}

\def\bbF{\mbox{\bbbold F}}

\def\bbM{\mbox{\bbbold M}}

\def\bbP{\mbox{\bbbold P}}

\def\bbR{\mbox{\bbbold R}}

\def\bbT{\mbox{\bbbold T}}

\def\cL{{\cal L}}

\newfont{\goth}{eufm10 scaled \magstep1}

\def\gs{\mbox{\goth s}}

\def\gu{\mbox{\goth u}}

\def\a{\alpha}
\def\b{\beta}
\def\C{\Gamma}
\def\d{\delta}\def\D{\Delta}
\def\e{\epsilon}

\def\h{\eta}

\def\l{\lambda}

\def\p{\pi}
\def\P{\Pi}

\def\s{\sigma}\def\S{\Sigma}

\def\be{\begin{equation}}\def\ee{\end{equation}}
\def\bea{\begin{eqnarray}}\def\eea{\end{eqnarray}}
\def\ba{\begin{array}}\def\ea{\end{array}}

\def\del{\partial}

\def\xz{\times}

\def\nab{\nabla}

\def\del{\partial}

{}

\def\bd{\begin{document}}
\def\ed{\end{document}}
\def\bea{\begin{eqnarray}}\def\barr{\begin{array}}\def\earr{\end{array}}
\def\eea{\end{eqnarray}}
\def\ft#1#2{{\textstyle{{\scriptstyle #1}\over {\scriptstyle #2}}}}
\def\fft#1#2{{#1 \over #2}}
\newcommand{\eq}[1]{(\ref{#1})}
\def\eqs#1#2{(\ref{#1}-\ref{#2})}
\def\det{{\rm det\,}}
\def\tr{{\rm tr}}

\begin{document}

\thispagestyle{empty}

\hfill{KCL-TH-98-62}

\hfill{hep-th/9812133}

\hfill{\today}

\vspace{20pt}

\begin{center}
{\Large{\bf On Harmonic Superspace}}
\vspace{30pt}

{P.S. Howe}
\vskip 1cm
{Department of Mathematics}
\vskip 1cm
{King's College, London}
\vspace{15pt}

\vspace{60pt}

{\bf Abstract}

\end{center}

A short survey of some aspects of harmonic superspace is given. In particular,
the $d=3, N=8$ scalar supermultiplet and the $d=6, N=(2,0)$ tensor multiplet
are described as analytic superfields in appropriately defined harmonic
superspaces.

{\vfill\leftline{}\vfill
\vskip	10pt
\footnoterule
\footnotesize{Contribution to the seminar ``Supersymmetries and Quantum
Symmetries'' held in Dubna, Russia, June 22-36, 1997 and dedicated to the
memory of Professor Victor I. Ogievetsky.}
\pagebreak
\setcounter{page}{1}

\section{Introduction}

Victor Ogievetsky made many significant contributions to theoretical physics
during a long and distinguished career often conducted in difficult
circumstances.
Perhaps his most interesting work, at least from this author's point of view,
was the development of a deeper understanding of the geometry of supersymmetry.
This began in the seventies with his and Sokatchev's interpretation of the
geometry of $N=1$ supergravity as the generalisation of chirality to the curved
superspace context. Some years later he and his group, now expanded to include
Galperin, Ivanov and Kalitzin as well as Sokatchev (GIKOS), succeeded in
finding the most useful generalisation of chirality to the case of flat space
$N=2$ supersymmetry, namely $N=2$ harmonic superspace. This new approach to the
geometry of supersymmetry has turned out to have wide applications to both
geometry and physics. In this note I shall briefly review flat harmonic
superspace, its relation to twistor theory, its generalisation to curved
superspace and its potentially interesting application to some on-shell
supermultiplets which arise naturally in string theory in the context of the
Maldacena conjecture.


\section{Harmonic superspace}

In extended supersymmetry flat harmonic superspaces \cite{gikos:N=2} are
superspaces of the form $M_H=M\xz \bbF$ where $M$ is the corresponding
Minkowski superspace and $\bbF$ is a coset space of the internal symmetry group
which is chosen to be a compact, complex manifold, in fact a flag manifold. For
example, in $D=4$, the internal symmetry group is $U(N)$ (or sometimes just
$SU(N)$), the relevant flags sequences $V_{k_1}\subset
V_{k_2}\subset....\subset\bbC^N$ of subspaces with dimension $k_i$, and the
flag manifold is the space whose points correspond to such flags. In
particular, the flags $V_p\subset V_{N-q}\subset\bbC^N,\ p+q\leq N$, define
$(N,p,q)$ harmonic superspace \cite{hh1} for which $\bbF=H\bsh U(N)$, the
isotropy group being $U(p)\xz U(N-(p+q))\xz U(q)$. This family generalises in a
natural way the harmonic superspaces for $N=2$ and $N=3$ supersymmetry
introduced by GIKOS, which are respectively $(2,1,1)$ \cite{gikos:N=2} and
$(3,1,1)$ \cite{gikos:N=3} harmonic superspace in the above notation.
\footnote{(3,2,1) superspace was first discussed in \cite{gikos:N=3ss}}

The harmonic superspace approach of GIKOS emphasises the group theoretic
aspects of fields defined on such spaces rather than the  holomorphic aspects.
For this reason it has become standard practice to work on the space $\hat
M_H=M\xz G$ (where $G=U(N)$ or whatever internal symmetry group is
appropriate). This is equivalent to working on $M_H$ provided that the fields
are restricted to being equivariant with respect to the subgroup $H$. Such a
field is a map $F:\hat M_H\rightarrow V$, where $V$ is a representation space
for the group $H$, such that $F(z,hu)=M(h)F(z,u)$, for all $u\in G,\ h\in H$
and $z\in M$, where $M(h)$ denotes the action of $H$ on $V$. An equivariant
field of this type defines in a natural way a section of a vector bundle $E$
over $M_H$ with typical fibre $V$ and the two types of object are in one-to one
correspondence.

In $D=4$ $(N,p,q)$ harmonic superspace $u$ is written in index notation as
$u_I{}^i$ where the $i$ index is acted on by $U(N)$ and the $I$ index by $H$,
and the inverse $u^{-1}$ is denoted $u_i{}^I$. The index $I$ splits under $H$
as $(r,R,r')$ where $r=1,\ldots p,\ R=p+1,\ldots N-q,\ r'=N-q + 1,\ldots N$.
The following derivatives can be defined on equivariant fields $F$,

\be
D_{\a r}=u_r{}^i D_{\a i};\qquad D_{\adt}^{r'}=u_i{}^{r'}\bar D_{\adt}^i
\ee

These derivatives anticommute and so allow the introduction of generalised
chiral fields, or Grassmann-analytic ($G$-analytic) fields, $F$, which satisfy

\be
D_{\a r}F=D_{\adt}^{r'}F=0
\ee

Moreover, for any fixed values of the $u's$ the derivatives $\{D_{\a
r},D_{\adt}^{r'}\}$ define a $CR$ structure on $M$; that is, they are basis
vectors for an involutive subundle $\bar K$ of the complexified tangent bundle
$T_c$ and their complex conjugates are linearly independent of them at any
point in $M$.\footnote{The r\^{o}le of $CR$ structures in harmonic superspace
was first emphasised by \cite{rs}.} The space $\bbF$ can thus be viewed as the
space of all $CR$ structures of this type on $M$. In addition, the derivatives
can be combined with a subset of the right-invariant vector fields $D_I{}^J$ on
$U(N)$ to define a $CR$ structure on $M_H$. The right-invariant vector fields
decompose under $H$ into the following subsets:

\be
D_I{}^J=\{ D_r{}^s,D_R{}^S,D_{r'}{}^{s'}\};\ \ \{D_r{}^S, D_r{}^{s'},
D_R{}^{s'}\};\ \ \{D_R{}^s, D_{r'}{}^s, D_{r'}{}^S \}
\ee

The first subset correponds to the isotropy group, the second to the components
of the usual $\bar\del$ operator on $\bbF$ and the third to its conjugate
$\del$. The $CR$ structure on $M_H$ is then determined by the derivatives
$\{D_{\a r}, D_{\adt}^{r'}, D_r{}^S, D_r{}^{s'}, D_R{}^{s}\}$. Since the
$D_I{}^J$ are characterised by

\be
D_I{}^Ju_K{}^k=\d_K{}^J u_I{}^k;\ D_I{}^Ju_k{}^K=-\d_I{}^K u_k{}^J
\ee

it is straightforward to verify that this set of derivatives does indeed
specify a CR structure. Finally, when $p=q$, there is a real structure on
harmonic superspace defined by $u\mapsto \e u$ where

\be
\e=\left(\barr{lll} 0 & 0 & 1_p \\ 0 & 1_{N-2p} & 0\\ -1_p & 0 & 0\earr\right)
\ee

This can be combined with complex conjugation to act on equivariant fields by

\be
F(z,u)\mapsto \widetilde{F(z,u)}=\overline{(F(z,\e u))}
\ee

For fields transforming under certain representations of the isotropy group it
is possible to demand that they be real with respect to this conjugation.


\section{Relation to twistor theory}

The above discussion is reminiscent of Euclidean twistor theory for $\bbR^4$
where $\bbC\bbP^1$ parametrises the space of (anti)-self-dual complex
structures on $\bbR^4$ and where the antipodal map of $\bbC\bbP^1$ defines the
real structure (as it does in (2,1,1) harmonic superspace) (see, for example
\cite{penrose,wardwells}). Indeed, particularly in cases where the appropriate
symmetry group is the superconformal group, flat-space harmonic superspace
theory can be viewed as an example of the group-theoretic approach to twistor
theory \cite{be}. The full power of this method only becomes apparent in
complexified spacetime (or superspace), in which case the generalised twistor
method starts with the selection of two parabolic subgroups $P_1, P_2$ of a
complex simple group $G$. Since $P_{12}\equiv P_1\cap P_2$ is also parabolic
there are three (complex) homogeneous spaces $\bbM_k=P_k\bsh G,\ k=1,2\ {\rm
or}\ 12$ which fit together into a double fibration
$\bbM_2\stackrel{\p_2}\leftarrow \bbM_{12}\stackrel{\p_1}\rightarrow \bbM_1$
such that each projection injects the fibres of the other one and thereby sets
up a correspondence between points of $m_2\in\bbM_2$ and subsets
$\p_1\circ\p^{-1}(m_2)\subset \bbM_1$ and vice versa. The basic idea of twistor
theory applied to physics is that one of the spaces, say $\bbM_1$, is spacetime
or Minkowski superspace and that information about field theory on (super)
Minkowski space can be encoded as holomorphic data on $\bbM_2$. $N=1$
supersymmetry in four dimensions provides an exceptional case where $\bbM_1$
and $\bbM_2$ are taken to be left and right chiral superspaces respectively and
$\bbM_{12}$ is super Minkowski space \cite{manin}.

In four dimensions the complex superconformal group is $SL(4|N)$. The parabolic
isotropy groups define flag supermanifolds where the flags are now made of
sequences of super subspaces of $N$-extended twistor space $\bbT_N=\bbC^{4|N}$,
and the superspaces of interest are constructed as certain open subsets of
these supermanifolds \cite{manin}, see also \cite{hh2}. (The flag
supermanifolds themselves have compact bodies). As an example consider $N=2$.
The group $SL(4|2)$ consists of supermatrices of the form

\be
g=\left(\barr{c|c} A & \C\\ \hline \D & B\earr\right)
\ee

where $A$ and $B$ are $4\xz 4$ and $2\xz 2$ even matrices, $\C$ and $\D$ are
$4\xz 2$ and $2\xz 4$ respectively odd matrices, and where ${\rm sdet} g=1$.
$N=2$ super Minkowski space is the space of flags of type (i.e. dimension)
$((2|0),(2|2))$ specified by the isotropy group consisting of matrices of the
form

\begin{equation}
 \left(\begin{array}{cccc|cc}
        \times & \times & & & &  \\
        \times & \times & & & &  \\
        \times & \times & \times & \times & \times &  \times \\
        \times & \times & \times & \times & \times &  \times \\ \hline
        \times & \times & & & \times &   \times \\
        \times & \times & & & \times &   \times
        \end{array} \right)
\end{equation}

where the crosses denote entries which are not necessarily zero. The body of
this superspace can be identified from the top left part of the diagram which
represents ordinary complex Minkowski space. Complex $N=2$ harmonic superspace
$\bbM_H$ is the space of flags of type $((2|2),(2|1),(2|2))\subset \bbT_2$
specified by the subgroup of matrices of the form

\begin{equation}
 \left( \begin{array}{cccc|cc}
        \times & \times & & & & \\
        \times & \times & & & & \\
        \times & \times & \times & \times & \times & \times \\
        \times & \times & \times & \times & \times & \times \\ \hline
        \times & \times & & & \times & \\
        \times & \times & & & \times & \times \end{array}
        \right)
\end{equation}

Note that this differs from super Minkowski space only in the bottom right-hand
corner. Thus, $N=2$ harmonic superspace has the same odd dimensionality as
$N=2$ super Minkowski space and its body is locally $\bbM\xz \bbC\bbP^1$. These
two superspaces fit into a double fibration with analytic superspace $\bbM_A$
which the space of flags of type $(2|1)\subset\bbT_2$. The isotropy group
consists of matrices of the form

\begin{equation}
\left( \begin{array}{cccc|cc}
                  \times & \times & & & \times & \\
                  \times & \times & & & \times & \\
                  \times & \times & \times & \times & \times & \times \\
                  \times & \times & \times & \times & \times & \times \\
                  \hline \times & \times & & & \times & \\
                  \times & \times & \times & \times & \times & \times
                  \end{array} \right)
\end{equation}

The odd dimensionality of analytic superspace is half that of harmonic
superspace but its body is the same. The double fibration is
$\bbM_A\stackrel{\p_2}\leftarrow\bbM_H\stackrel{\p_1}\rightarrow \bbM$. Points
of $\bbM$ correspond to twistor lines, copies of the fibres of $\p_1$, in
$\bbM_A$, while points of $\bbM_A$ parametrise certain planes of dimension
$(0|4)\subset\bbM$, and these are copies of the fibres of $\p_2$. Note that
analytic superspace is essentially a complex space; it does not exist as a
coset space of the real $N=2$ superconformal group $SU(2,2|2)$. Furthermore, it
is not a subspace of complex harmonic superspace, but rather a quotient space
defined by the projection $\p_2$. The antipodal map on $\bbC\bbP^1$ defines an
anti-holomorphic involution of $\bbM_A$ and the real twistor lines in $\bbM_A$
are parametrised by points in real Minkowski superspace. When super Minkowski
space is taken to be real, i.e. when $\bbM$ is replaced by $M$, real harmonic
superspace $M_H=\p_1{}^{-1}(M)$ and $\p_2$ embeds $M_H$ as a submanifold of
$\bbM_A$. In the real case, therefore, the double fibration can be replaced by
the single fibration $M_H\rightarrow M$ while holomorphic fields on $\bbM_A$
are replaced by $CR$-analytic fields on $M_H$ \cite{hh2}.

The above discussion generalises in a natural way to $(N,p,q)$ harmonic
superspace \cite{hh1,hh2}. An entirely different family of supermanifolds can
be constructed by using isotropy groups which are completely filled in the
bottom right corner, but which have different structures to Minkowski space in
the top left corner. Such superspaces, which are usually referred to as
supertwistor spaces and which have conventional twistor spaces as bodies, have
also been used in the context of supersymmetric field theory (see, for example,
\cite{witten,manin}).


\section{Curved harmonic superspace}

The notion of harmonic superspace generalises to curved superspace relatively
straightforwardly \cite{gikos:N=2sg,gikos:N=2csg,hh1}. Perhaps the simplest
geometries that can be discussed in this context are $D=4$ superconformal
geometries ($N\leq 4$) \cite{h1}. A superconformal structure on a real $(4|4N)$
supermanifold is a choice of odd tangent bundle $F$ (rank $(0|4N)$) such that
the Frobenius tensor, defined by evaluating the commutator of two odd vector
fields (sections of $F$) modulo $F$ has the following components with respect
to bases $\{E_{\a i},\bar E_{\adt}^i\}$ of $F$ and $\{E^a\}$ for $B^*$ where
$B=T/F$:

\be
T_{\a i\b j}{}^c=T^{ij}_{\adt\bdt}{}^c=0;\qquad
T_{\a i \bdt}^{\phantom{\a i} j c}=-i\d_i{}^j(\s^c)_{\a\bdt}
\ee

The Frobenius tensor is just the dimension zero component of the usual torsion
tensor, although it is not necessary to introduce a connection to define it.
With the components satisfying (11) above it is invariant under the group
$\bbR^+\xz SL(2,\bbC)\cdot U(N)$ acting on $F$. The above constraints can be
shown to lead to the usual constraints of off-shell conformal supergravity if
standard conventional choices are made for the even tangent bundle $B$ and for
the $SL(2,\bbC)\cdot U(N)$ connection \cite{h1}. For $N=4$ an additional
constraint must be imposed arising from the fact that the true internal gauge
symmetry group in this case is $SU(4)$.

In the curved case the space $\hat M_H$ is the (perhaps locally defined)
principal $U(N)$ bundle associated with the $U(N)$ factor of the structure
group of $F$, and harmonic superspace $M_H$ itself is then given as $H\bsh\hat
M_H$ where $H$ is the relevant isotropy group, so that $\hat M_H$ is also a
principal $H$-bundle over $M_H$.

For $N=1$ harmonic superspace is not necessary and the constraints (11) above
imply that a real $(4|4)$ supermanifold has a superconformal structure if and
only if it has a $CR$ structure with involutive $CR$ bundle of rank $(0|2)$
such that the Frobenius tensor completely specifies $B$. This is
Ogievetsky-Sokatchev supergeometry \cite{os}.

For $N=2$, the appropriate harmonic superspace is $(2,1,1)$ superspace
\cite{gikos:N=2csg}. There is a natural bundle $\bar K$ on $M_H$ defined by the
set of basis vector fields $D_1{}^2$ and the horizontal lifts of the odd basis
vector fields on $M$ projected in the $(\a 1)$ and $\left(\barr{c}
2\\\adt\earr\right)$ directions. To define the latter it is necessary to
introduce a $U(2)$ connection. This bundle defines a $CR$ structure if and only
if the contraints of $N=2$ superconformal geometry are satisfied \cite{hh1}.
Note, however, that there is no gauge-invariant notion of $G$-analyticity. This
problem was circumvented in \cite{gikos:N=2csg} by allowing $u_1{}^i$ and
$u_2{}^i$ to be independent, but the geometrical status of this strategy could
perhaps be made more precise.

For $N=3$ in $(3,1,1)$ harmonic superspace a similar $CR$ structure is
compatible with the constraints of conformal supergravity but does not imply
them \footnote{This point has been observed by others (E. Sokatchev, private
communication).}, although demanding that the lift of the Frobenius tensor to
$M_H$ should vanish along the $(\a 1)$ and $\left(\barr{c} 3\\\adt\earr\right)$
directions does. In addition, at the linearised level, the conformal
supergravity potential in harmonic superspace is a dimension $-2$ $G$-analytic
field $V_1{}^3$. This field can be interpreted as a 3-form along the
antiholomorphic fibre directions, but a clear geometrical understanding is
lacking at present.

Finally, for $N=4$, it is possible to use $(4,2,2)$ harmonic superspace in
which case the natural $CR$ structure is equivalent to the constraints given
above in (11). However, as noted previously, an additional constraint has to be
imposed and this should eventually lead to the construction of a prepotential
which is known to be a $G$-analytic 4-form along the antiholomorphic fibre
directions in the linearised theory \cite{hh1}. The case of $N=4$ conformal
supergravity is particularly interesting in view of its relevance to the
Maldacena conjecture.


\section{Superconformal fields}

In many supersymmetric theories for which there is no known off-shell
formulation the fundamental field may be considered to be a field strength
tensor, and in many cases this field turns out to have a simple description as
a single-component $CR$-analytic field $W$ on an appropriate harmonic
superspace. The hope is that this point of view may be useful in constructing,
or attempting to construct, non-perturbative correlation functions of
gauge-invariant composite operators made from powers of $W$ by exploiting
$CR$-analyticity, particularly when the theory under consideration is
superconformal. This approach to quantum supersymmetry has been studied
extensively in the case of $N=4$ Yang-Mills theory (see, for example,
\cite{hw1}, \cite{hw2} and \cite{hsw}) and is currently highly topical in that
it is directly relevant to the Maldacena conjecture which relates supergravity
theories on anti-de Sitter backgrounds to conformal field theories on the
boundary, i.e. (super) Minkowski space \cite{m}.

The multiplets to be discussed here are matter (i.e. non-gravitational)
multiplets with the maximal number of supersymmetries, sixteen. The basic
multiplets are the $D=4, N=4$ Yang-Mills field strength multiplet, the $D=6,
(2,0)$ tensor multiplet and the $D=3, N=8$ scalar multiplet. They are relevant
to IIB string theory on $AdS_5\xz S^5$, M-theory on $AdS_7\xz S^4$ and M-theory
on $AdS_4\xz S^7$ respectively, although the latter two cases are less well
understood as these multiplets are only known at the non-interacting level.
Nevertheless they are extremely interesting multiplets, corresponding as they
do to the multiplets of the M-theory 5-brane and 2-brane respectively.

\subsection{$D=4, N=4$ Yang-Mills}

The field strength superfield $W_{ij}=-W_{ij}$ in super Minkowski space
transforms according to the six-dimensional representation of the internal
symmetry group $SU(4)$ and satisfies the following conditions:

\bea
\nab_{\a i} W_{jk}& =&\nab_{\a [i} W_{jk]}\\
\bar\nab_{\adt}^i W_{jk}& =& -{2\over3}\d_{[j}{}^i\nab_{\adt}^l W_{k]l}\\
W^{*ij}&=&{1\over2}\e^{ijkl} W_{kl}
\eea

The appropriate superspace in this case is $(4,2,2)$ harmonic superpace
specified by the internal flag manifold $\bbF=S(U(2)\xz U(2))\bsh SU(4)$. An
element $u\in SU(4)$ is written $u_I{}^i=(u_r{}^i,u_{r'}{}^i),\ r=1,2;\
r'=3,4$. The right-invariant derivatives on $SU(4)$ are $\{D_r{}^s,
D_{r'}{}^{s'},D_o; D_r{}^{s'}; D_{r'}{}^s\}$, where $\{D_r{}^s,
D_{r'}{}^{s'},D_o\}$ correspond to the isotropy algebra
$\gs\gu(2)\oplus\gs\gu(2)\oplus\gu(1)$, $D_{r}{}^{s'}$ corresponds to the
$\bar\del$ operator on $\bbF$ and $D_{r'}{}^s$ to its conjugate $\del$. The
normalisation is

\be
D_o u_r{}^i= {1\over2}u_r{}^i;\ D_o u_{r'}{}^i= -{1\over2}u_{r'}{}^i
\ee

The $CR$-structure on harmonic superspace is specified by the set of
derivatives $\{D_{\a r},D_{\adt}^{r'},D_r{}^{s'}\}$.

The claim is that $W_{ij}$ is equivalent to a charge 1 field $W$ on $M_H$ which
is covariantly $G$-analytic and ordinarily $\bbF$-analytic; it is also real
with respect to the real structure discussed in section 2 above, where
covariantly $G$-analytic means that

\be
\nab_{\a r} W=\nab_{\adt}^{r'} W=0
\ee

with $\nab_{\a r}=u_r{}^i\nab_{\a i}$, etc. This claim is easily verified
\cite{hh1}.

The gauge-invariant operators $A_q=\tr (W^q)$ are $G$-analytic in the usual
sense and hence $CR$-analytic. These are the conformal fields of interest in
the Maldacena conjecture. Since they are in short representations of
$SU(2,2|4)$, the integer $q$ cannot be affected by quantum corrections and so,
since this integer determines the dimension, they do not have anomalous
dimensions. This family of operators was introduced in \cite{hw1} and it has
been shown that it is in one-to-one correspondence with the Kaluza-Klein
spectrum of IIB supergravity on $AdS_5\xz S^5$ \cite{af}. In particular, the
family of operators includes the energy-momentum tensor $T=A_2$ \cite{hst}. In
complex spacetime $A_q$ becomes a holomorphic section of the $q$th power of a
certain homogeneous line bundle $\cL$ over analytic superspace. This superspace
is in fact the super-Grassmannian whose points are $(2|2)$ planes in
$\bbT_4=\bbC^{4|4}$.

\subsection{The $D=6, (2,0)$ tensor multipet}

In six-dimensional $(n,0)$ supersymmetry the internal symmetry group is $Sp(n)$
and the spinors are taken to be symplectic Majorana-Weyl. The (2,0) tensor
multiplet is a scalar superfield $W_{ij}$ which transforms according to the
real 5-dimensional representation of $Sp(2)$ (vector representation of
$SO(5)$). It satisfies the constraints \cite{hsit}

\bea
W_{ij}&=&-W_{ji};\qquad \h^{ij}W_{ij}=0\\
\bar W^{ij}&=& \h^{ik}\h^{jl} W_{kl}\\
D_{\a i} W_{jk}&=& 2\h_{ij}\l_{\a k}-2\h_{ik} \l_{\a j}+\h_{jk}\l_{\a i}
\eea

where $\l_{\a i}= {1\over5}\h^{jk} D_{\a j} W_{ki}$, and where $\h_{ij}$ is the
antisymmetric $Sp(2)$ invariant tensor. The components of $W_{ij}$ consist of
five scalars, four spinors and a self-dual three-form field strength tensor.

The appropriate harmonic superspace in this case is $M_H=M\xz \bbF$ where
$\bbF=U(2)\bsh Sp(2)$ \cite{hprep}. Elements $u$ of the group $Sp(2)$ are
written $u_I{}^i=(u_r{}{}^i,u_{r'}{}^i),\ r=1,2;\ r'=3,4.$ In addition to being
unitary the matrices $u$ also satisfy $u_I{}^i u_J{}^j \h_{ij}=\h_{IJ}$. The
matrix $\h$ has components

\be
\h_{rs}=\h_{rs'}=0;\ \h_{rs'}=-\h_{s'r}= \d_{rs'}
\ee

The right-invariant vector fields $\hat D_{IJ}$ on the group are real,
symmetric and satisfy

\be
\hat D_{IJ}u_K{}^i=-\h_{IK} u_J{}^i-\h_{JK} u_{I}{}^i
\ee

The derivatives split into the isotropy group derivatives $\hat D_{rs'}$ and
the coset derivatives $\hat D_{rs}\equiv D_{rs}$ and $\hat D_{r's'}\equiv
D_{r's'}$. The isotropy group derivatives may be separated into a $\gu(1)$
derivative $D_o={1\over2}\h^{rs'}D_{rs'}$ and $\gs\gu(2)$ derivatives
$D_{rs'}=\hat D_{rs'}-\h_{rs'}D_o$. On $u$, $D_o u_r{}^i= u_r{}^i; D_o
u_{r'}{}^i=-u_{r'}{}^i$. The element $u_{r}{}^i$ transforms as a doublet under
$\gs\gu(2)$ while $u_{r'}{}^i$ transforms as the conjugate doublet.

The $CR$ structure on $M_H$ is specified by the derivatives $D_{\a r}=u_r{}^i
D_{\a i}$ and $D_{rs}$, the latter corresponding to the components of
$\bar\del$ on $\bbF$ as usual.

In a very similar manner to the $D=4, N=4$ Yang-Mills case discussed above it
can easily be shown that the field $W_{ij}$ is equivalent to a single-component
$CR$-analytic field $W$ on $M_H$ with $W={1\over2}\e^{rs}u_r{}^i u_s{}^j
W_{ij}$. It is also real with respect to a suitably defined real structure.
Although there is no known non-Abelian version of this multiplet it is still
possible to use it to discuss conformal field theory in the abstract
\cite{sei}. The powers of $W$ give conformal fields in short representations,
in particular $W^2=T$ is again the energy-momentum tensor multiplet (conformal
supercurrent) \cite{hsit}; a few of them were listed in \cite{roz}. It is
likely that the complete set of such fields corresponds to the Kaluza-Klein
spectrum of eleven-dimensional supergravity on $AdS_7\xz S^4$.

\subsection{D=3, N=8 scalar multiplet}

The scalar multiplet in three dimensions is slightly trickier to describe than
the preceeding two cases because the internal symmetry group is $SO(8)$ and it
is necessary to use both spinor and vector representations of the group to
describe the multiplet. Nevertheless it can be done and the result is simply
stated: the $D=3, N=8$ scalar multiplet can be represented by a
single-component $CR$-analytic superfield defined on the harmonic superspace
$M_H=M\xz U(4)\bsh Spin(8)$ \cite{hprep}.

The multiplet is defined in $D=3, N=8$ Minkowski space by the equation

\be
D_A W_I = (\S_I)_{AA'} \l^{A'}
\ee

where $I,A$ and $A'$ transform according to  the vector, spinor and primed
spinor representations of $Spin(8)$, $\S$ denotes the spin matrices and where
the $D=3$ spinor indices on both $D$ and $\l$ have been suppressed. The spinor
representations decompose into 4 and $\bar 4$ representations under $SU(4)$
with particular charges under $U(1)$ while the vector representation decomposes
into the six-dimensional representation of $SU(4)$, neutral under $U(1)$ and a
charged singlet and its conjugate under $U(1)$. That is $v_I\rightarrow
(v_{++},v_{--},v_i),\ i=1\ldots 6$, where $v_{++}={1\over2}(v_7-iv_8),
v_{--}=\bar{v}_{++}$, while for a $D=8$ Majorana spinor $\Psi$, with

\be
\Psi =\left(\barr{c} \psi\\\chi\earr\right)
\ee

the indices are split as follows,

\be
\psi_A\rightarrow \left(\barr{c} \psi_{\a +}\\
\bar\psi^{\a}_{-}\earr\right)\qquad
\chi^{A'}\rightarrow\left( \barr{c} \chi_{\a
-}\\\-\bar\chi^{\a}_{+}\earr\right)
\ee

The lower (upper) $\a$ indices correspond to the $4 (\bar 4)$ representations
of $SU(4)$.

The right-invariant derivatives on the group split into the isotropy
derivatives $\{D_{ij}, D_o\}$ and the coset derivatives $\{D_{++i}, D_{--i}\}$.
The $CR$ structure is specified by the derivatives $D_{\a +}=u_{\a +}{}^A D_A$
and $D_{++i}$. The field $W_{++}=u_{++}{}^I W_I$ is easily seen to be
$CR$-analytic, as well as being real with respect to an appropriate real
structure.

In both of the latter two cases the superconformal group in complex spacetime
is $OSp(8|2)$ and the analytic superspaces are the same, although the line
bundles for the field strengths are different. However, the underlying twistor
spaces are $\bbC^{4|8}$ for $D=3$ and $\bbC^{8|4}$ for $D=6$, the two being
related by the Grassmann parity flip operation $\P$. This may be a reflection
of ``electromagnetic'' duality in $D=11$; the 2-brane couples to the 4-form
field strength of $D=11$ supergravity while the 5-brane couples to its 7-form
dual. In case (a) on the other hand, the D3-brane is self-dual, as is $D=4,
N=4$ twistor space under $\P$.

\end{document}